\documentclass[conference]{IEEEtran}

\usepackage{microtype}                 
\usepackage{textcomp}                  
\usepackage{mathptmx}                  
\usepackage{times}                     
\usepackage{cite}                      
\usepackage{tabu}                      
\usepackage{booktabs}  
\usepackage{graphicx}
\graphicspath{ {pictures/} }

\usepackage{algorithm}
\usepackage[noend]{algpseudocode}

\makeatletter
\def\BState{\State\hskip-\ALG@thistlm}
\makeatother

\usepackage{varwidth}
\usepackage{tabularx}
\usepackage{pgfplots}
\pgfplotsset{width=8.2cm,compat=1.9}
\usepgfplotslibrary{external}
\usepackage[margin=0.7in]{geometry}

\begin{document}
\title{TRUFL: Distributed Trust Management framework in SDN}

    \author{
    \IEEEauthorblockN{Ankur Chowdhary$^\ast$, Dijiang Huang$^\ast$, Adel Alshamrani$\dagger$,\\ Myong Kang$\ddagger$, Anya Kim$\ddagger$ and Alexander Velazquez$\ddagger$}
    \IEEEauthorblockA{Arizona State University$^\ast$, University of Jeddah$\dagger$, Naval Research Lab$\ddagger$
        \\\{achaud16, dijiang\}@asu.edu, asalshamrani@uj.edu.sa,\\ \{myong.kang, anya.kim, alexander.velazquez\}@nrl.navy.mil}

        
}


%


\maketitle

\begin{abstract}
Software Defined Networking (SDN) has emerged as a revolutionary paradigm to manage cloud infrastructure. SDN lacks scalable trust setup and verification mechanism between Data Plane-Control Plane elements, Control Plane elements, and Control Plane-Application Plane. Trust management schemes like Public Key Infrastructure (PKI) used currently in SDN are slow for trust establishment in a larger cloud environment. We propose a distributed trust mechanism - TRUFL to establish and verify trust in SDN. The distributed framework utilizes parallelism in trust management, in effect faster transfer rates and reduced latency compared to centralized trust management. The TRUFL framework scales well with the number of OpenFlow rules when compared to existing research works.

 



\end{abstract}

%
\IEEEpeerreviewmaketitle


\section{Introduction}
The cloud networks are often distributed geographically but for ease of management, they have built-in separation of concern mechanism known as multi-tenancy. SDN has been utilized by many cloud providers recently for network orchestration. Although SDN framework provides some security applications, it was not created with security as end-goal in mind.

SDN switches are known to cache the flow entries for flow rules that have been installed in switch flow tables and only new flows are sent to the controller. There are several security attack vectors such as; a) Forged traffic flows that can affect communication between the control plane and data plane b) Lack of mechanisms to ensure trust between the control plane and management plane applications~\cite{lim2015security}. A malicious insider may take advantage of this fact and fill up the switch buffer in effect creating a switch Denial of Service (DoS) attack. Another scenario can be to generate a new variation of flow rules and mounting Distributed Denial of Service (DDoS) attack on the controller. OpenFlow is one of the most popular protocols, which allows the incorporation of SDN capabilities in the cloud network. OpenFlow switches consist of flow tables, which run at line rate to allow traffic between data plane devices.

\textbf{Scalability Issue in Centralized Trust Management:} Public Key Infrastructure (PKI) standard based on ITU-T recommendation X.509 supports the identification of trusted users in the system, generation and exchange for the secured key over the network and verification of authenticity of users using a digital signature. PKI model based on DNS has been proposed by IETF~\cite{sdnpki}, where each tenant controls its operation zone and operators or admin can define access control template for tenants. On an average 3-4 messages are exchanged between Certificate Authority (CA) client and server for key exchange and 2-3 additional messages for digital certificate verification. If we consider setup and verification time to be around 0.1s for one client, the overall procedure will take several minutes for a cloud network with 10000 hosts. 

Additionally, applications with heavy traffic bursts can lead to a slow performance in peak traffic situations because context switch between user and kernel space is a costly operation. DPDK~\cite{dpdk} can be used to eliminate the context switch above by using specialized drivers known as Poll Mode Drivers (PMDs). The packets can be directly transferred between user space and the network interface driver using PMD.

OpenStack based virtual infrastructure framework based on Open vSwitch can be optimized to improve performance and provide low-latency in packet processing using multi-threading offered DPDK. The results discussed in DPDK driver show packet processing performance enhancement by 2.5x times as discussed in report~\cite{inteldpdk}. 

If a trust management system like PKI is deployed in the context of SDN infrastructure, the validation of traffic at data plane and control plane level would be too slow for a large multi-tenant cloud environment.  We leverage the innovative approach of using parallelized trust management to optimize the setup and verification of trust management framework in an OpenStack cloud network. Key contributions of this research work are as follows:

\begin{enumerate}
\item Trust management framework in SDN to prevent malicious insiders from mounting security attacks like denial of service attacks (DoS). 
\item Distributed setup and verification of trust in SDN environment to reduce application overhead of PKI.
\item The forwarding latency of TRUFL is lower than exiting research works NetSyn~\cite{mcclurg2015efficient}, SDPA~\cite{zhu2015sdpa} and Hassel~\cite{kazemian2012header}.
\end{enumerate}
Although Intel and 6WIND~\cite{6wind} have been using DPDK with Open vSwitch to optimize packet processing, to the best of our knowledge, no one has used DPDK for security provisioning in a multi-tenant cloud network.

\section{Background}
SDN supports different protocols, third-party applications, and controllers. SDN can help an admin to centralize command and control of the cloud environment, a big concern is a trust between various components in SDN itself. There can be several ways in which trust can be violated by an attacker. A rogue insider can add a fake switch, additional host nodes which are not part of the SDN environment to achieve desired communication.

It can be quite hard to detect trust violations in the SDN framework. We need to ensure that flow rules across the infrastructure are compliant with high Service-Level Agreements (SLAs). If we utilize existing public key based trust management systems in SDN, scalability will be a major concern for large multi-tenant cloud networks.

\subsection{SDN and OpenFlow}

\textbf{OpenFlow Rule:} A flow table F of an OpenFlow switch, can have rules, $\{r_1, r_2,.., r_n\}$ Each rule consists of layer 2-4 packet header fields, protocol (TCP/UDP/FTP), action-set associated with the rule, rule priority, and statistics. We define the flow rule using tuple $r_i$ = ($p_i$, $\rho_i$, $h_i$, $a_i$, $s_i$), where a) $p_i$ denotes rule priority, b) $\rho_i$ denotes the protocol of the incoming traffic (TCP/UDP) c) $h_i$ depicts the packet header, d) $a_i$ is the action associated with the rule, e) $s_i$ represents the statistics associated with the rule. 
    
The flow rule header space $h_i$, consists of physical port of incoming traffic $\delta_i$, source and destination hardware address, i.e., ${\alpha_s}_i, {\alpha_d}_i$, source and destination IP address, ${\beta_s}_i, {\beta_d}_i$, source and destination port address, ${\gamma_s}_i, {\gamma_d}_i$. Packet header can be defined by the tuple $h_i$ = ($\delta_i$, ${\alpha_s}_i, {\alpha_d}_i$, ${\beta_s}_i, {\beta_d}_i$, ${\gamma_s}_i, {\gamma_d}_i$). Rule statistics $s_i$, comprises of both flow duration and number of packets/bytes for each flow rule $s_i = (d_i, b_i)$.

\subsection{PKI Model and Components}
PKI model we use mainly consists of the following components:-
\begin{enumerate}
    \item Certificate Authority (CA) is responsible for issuing and revoking certificates.
    \item Certificate Signing Request (CSR) Request for a certificate containing public key and ID to be certified.
    \item Certificate - public key and ID bound by a CA signature.
    \item Certificate Revocation List (CRL) - list of revoked certificates, issues by CA from time to time. 
    \item Digital Signature - used to verify the identity of clients.
\end{enumerate}
We have focused our attention on CA, CSR and digital signatures in this work. We plan to discuss CRL in SDN as part of future work. Other key terms used in this work have been defined in OpenSSL document~\cite{opensslpki}. We have omitted those definitions in text for sake of brevity. 

\subsection{Open vSwitch with DPDK - PKI Setup}
\begin{figure}[!ht]
    \centering
    \includegraphics[width=0.5\textwidth]{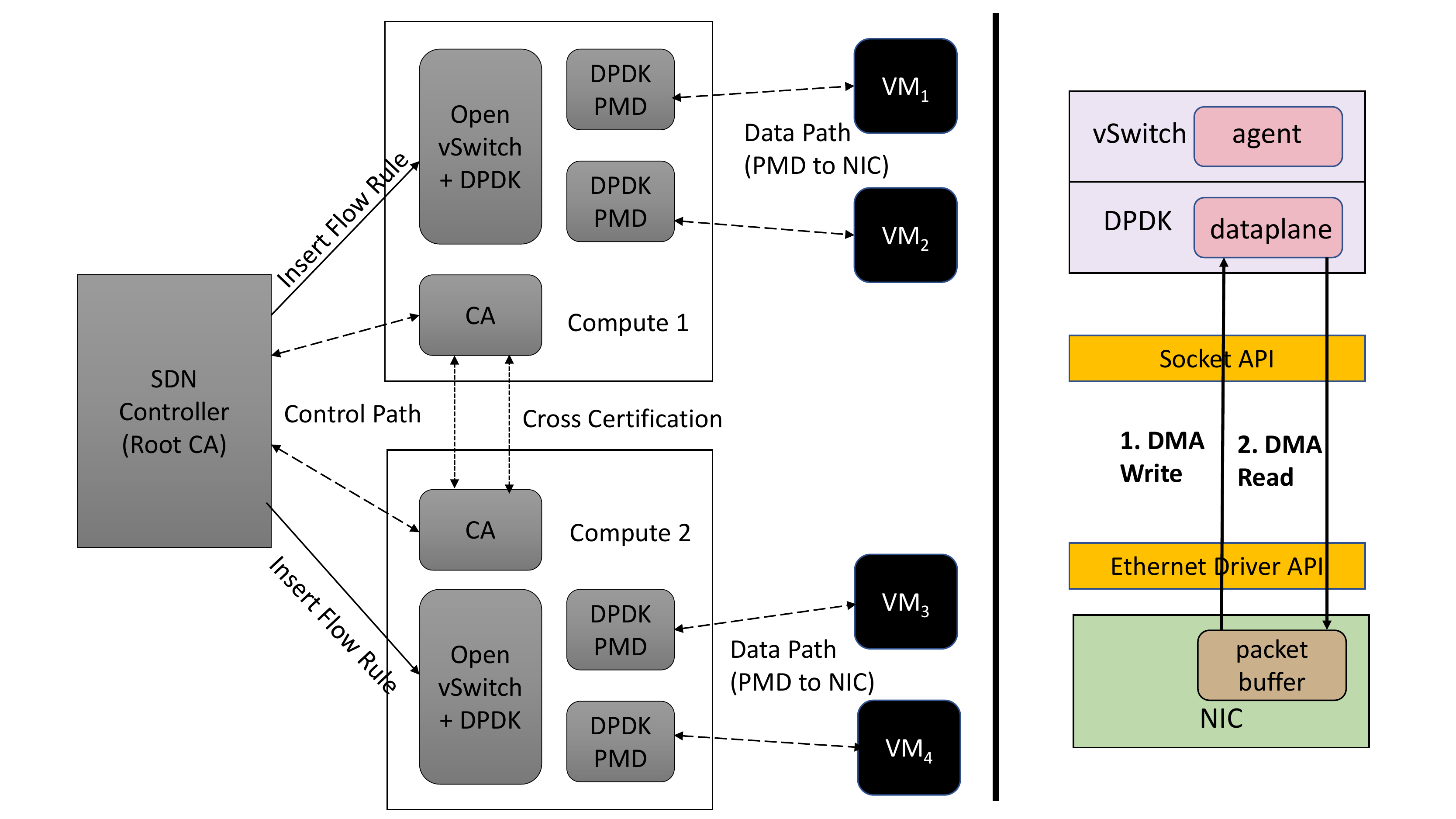}
    \caption{Fast Packet Processing and Trust Setup with DPDK}
    \label{fig:pmd}
\end{figure}

As shown in Figure~\ref{fig:pmd}, Poll Mode Driver (PMD) can access the network interface card of connected devices directly without any kernel level system call. This saves significant time in terms of context-switch. The input packets are received directly using Direct Memory Access (DMA) by polling Network Interface Card (NIC's) RX buffer using PMD receive API. Similarly, PMD transmits API is used to place output packets on the TX buffer of NIC.

We have used PMD's parallelized mode to allocate one core on a physical server hosting Open vSwitch (OVS) for each guest machine connected to Open vSwitch as a port. As can be seen in the figure, we use parallel processing to setup PKI infrastructure for each VM connected to Open vSwitch. The Open vSwitch acts as a CA server and connected hosts are acting as CA clients. PMD parallelization is used for PKI setup and verification.

We achieved fast PKI setup using a) DPDK based fast packet processing due to user-space operations based on DMA (no context switch) b) PMD based parallelized key exchange and verification for connected hosts. 
\section{Related Work}

There has been a good deal of work in identifying security issues in SDN. Most of the work at present try to address security issues from the flow rule violation aspect. There is, however a limited focus on securing the overall framework itself.

\textbf{Trust Management in SDN:}
FortNOX~\cite{porras2012security} is a security enforcement kernel that ensures role-based authorization and security policy enforcement in the SDN network. The authors describe different levels of security authorization i.e- administrator, security applications, normal OF applications. The flow rules are signed by the respective authorization role using the digital signature scheme. The flow rules of higher authorization level e.g., admin can override flow rule signed by lower authorization role e.g., OF appreciation. Signing and tracking of every new flow rule can, however, introduce latency in the network. We have therefore used PKI to verify the particular SDN elements and treat flow rule violation part separately in our current work.

Authors in~\cite{kreutz2013towards} discuss the new fault and attack plane capabilities that are introduced by SDN infrastructure because of centralized command and control which makes it easier for attackers to exploit infrastructure. The paper highlights seven major threat vectors including forged traffic flows, lack of trust between controller and management plane applications, attack on control plane communications. The paper recommends maintaining a whitelist of trusted devices, autonomic trust management and security domains to provide trust. We have covered most of these desired features in current work via better security design, PKI infrastructure and flow rule conflict checking mechanism in a cloud network.

\textbf{Distributed SDN Frameworks:}
Onix \cite{koponen2010onix} facilitates distributed control in SDN by providing each instance of the distributed controller access to holistic network state information through an API. HyperFlow \cite{tootoonchian2010hyperflow} synchronizes the network state among the distributed controller instances while making them believe that they have control over the entire network. Attack graph based, scalable security solution has been discussed by Chowdhary \textit{et al}~\cite{chowdhary2016sdn}. The paper~\cite{betge2015trust}  uses architecture to support mapping between management plane application and several controllers to tackle the issue of trust in SDN controllers. The separate execution environment is maintained for each controller, and incoming requests from different controllers are compared for consistency before flow rules are installed in OpenFlow tables. 


One major security~\cite{shin2013fresco} issue with SDN is controller acting as a single point of failure. FRESCO framework has introduced modular decomposition of control plane logic into several modules. These APIs are used for data sharing and event triggering between these modules. This approach will help improve the scalability and robustness of the system. We have implemented similar logic in application plane design where CAServer and Flow rule conflict checking modules~\cite{pisharody2016conflicts},~\cite{huang2018software} are separate and operate independently of each other. Rendezvous-based trust propagation has been discussed by authors~\cite{cheng2011rendezvous}. A similar model can also be incorporated for trust computation into SDN environment where the destination node can compute the trust of the source node via a recommendation of intermediate switches along the path.

\section{System Architecture}

\begin{figure}[!ht]
    \centering
    \includegraphics[width=0.5\textwidth]{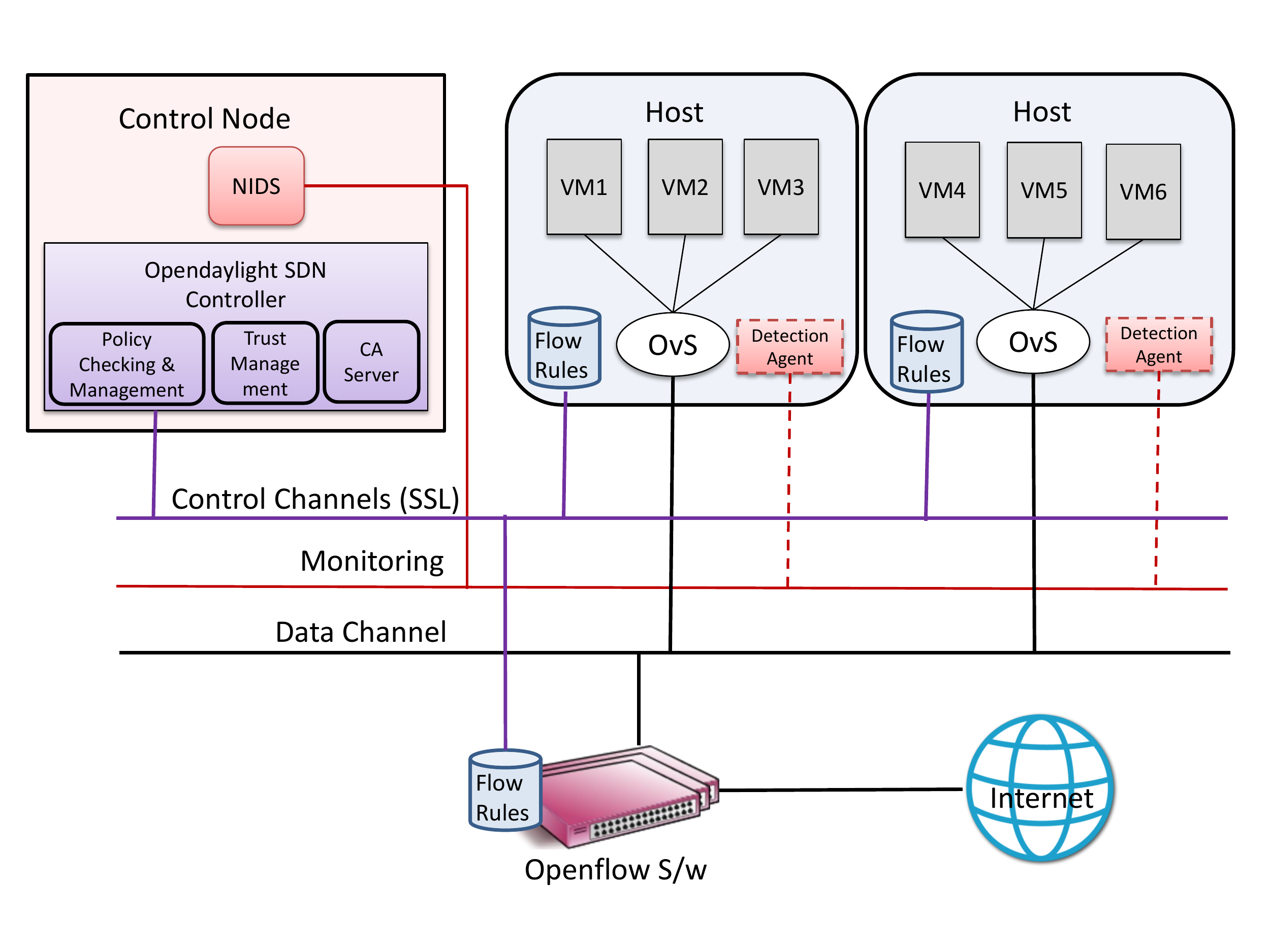}
    \caption{System Architecture}
    \label{fig:sysarch}
\end{figure}

We consider a multi-tenant cloud architecture. Each tenant will have a detection agent that is responsible for performing intrusion detection. Figure~\ref{fig:sysarch} shows the OpenStack based cloud framework where the VM's are spanned across two compute nodes. A centralized controller is responsible for controlling and coordinating networking and storage elements. 

Our SDN controller is part of the Control Node of OpenStack. The Opendaylight controller has several applications as part of a management plane, e.g., Flow rule conflict checker, one or more Certificate Authority (CA) to enforce trust between different compute node elements. The controller also comprises of other applications such as Network-based Intrusion Detection System (NIDS) which mirrors data plane traffic from different data plane VM's using port mirroring techniques. The NIDS uses neutron (network manager for OpenStack) based APIs to provide NIDS functionality. 

We primarily consider components which are part of the SDN controller to check trust violations by misconfiguration of flow rules or by the presence of rogue control and data plane elements in this work. SDN controller makes use of Secured Socket Layer (SSL) to communicate with Control Plane elements (Open vSwitch) on each Compute Node. The system is connected to the physical switch via an SSL connection as shown in Figure~\ref{fig:sysarch}.

\subsection{Threat Model}
We consider the threat model based on two different variants of attacks; 
a) Rogue insider is trying to send the malicious traffic flow to switch.
b) Fake switch trying to insert flow rules in the switch using elevated privileges. We have not considered a malicious outsider problem in this work because the traffic from malicious outsider has to pass through additional layers of a firewall to establish communication with high-value nodes inside the network. Trust violation in such a case is quite challenging for a malicious outsider. We have rather considered the case where some node within a trusted intranet environment has been compromised (rogue node) or some disgruntled employee is trying to exploit the trust framework in the network (insider threat).

\begin{figure}[!ht]
    \centering
    \includegraphics[width=0.5\textwidth]{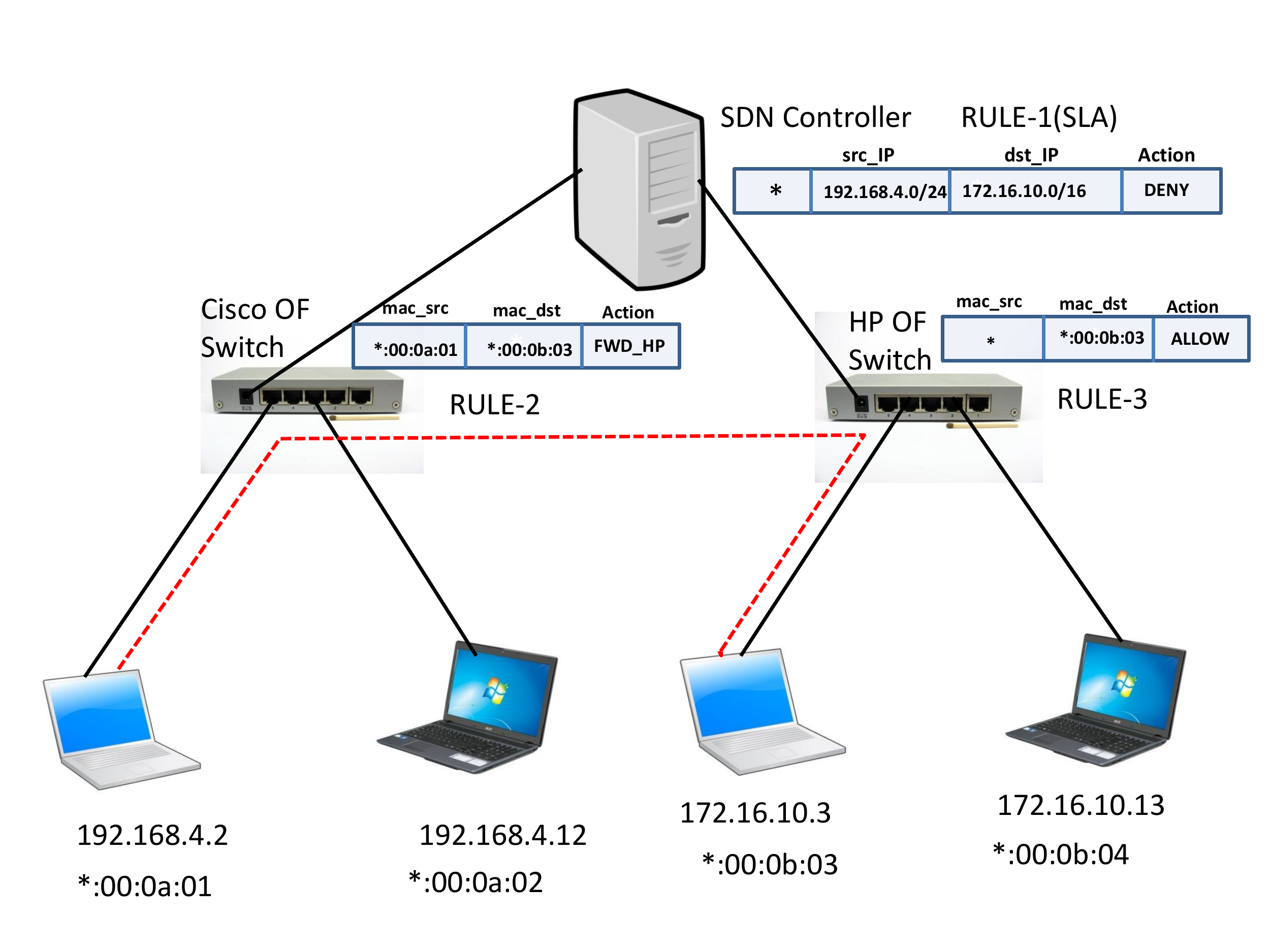}
    \caption{Motivating Example for Trust Violation}
    \label{fig:trust1}
\end{figure}

We consider an architecture similar to the setup of hosts and switches in a data-centric network. Typically, there are multiple hosts connected by one or more switches. We consider Cisco based OF Switch connected to two hosts and HP based OF switch connected to two hosts.

The high level SLA defined by controller blocks traffic flow between sub-networks \textit{192.168.4.0/24} and \textit{172.16.10.0/16} as depicted in Figure~\ref{fig:trust1}. There can be several ways in which a rogue insider or a compromised switch can bypass this SLA to achieve communication between host \textit{192.168.4.2} and \textit{172.16.10.3}. 

\textbf{Rogue Insider Attack:} 
We consider a rogue insider trying to bypass flow rule policy which denies his connection over layer 3 (IP traffic) to another host. The attacker can mount ARP cache poisoning or man-in-the-middle attack to spoof all traffic from the victim's machine at layer 2 using this attack. 

\begin{equation}
*|192.168.4.0/16| 172.16.10.0/16| DROP (SLA)
\end{equation}
\begin{equation}
*:00:0a:01| *:00:0b:03 |FWD\_HP (Switch1) \\
\end{equation}
\begin{equation}
*:*:*| *:00:0b:03 |ALLOW           (Switch2) \\
\end{equation}
The rule 1 clearly denies any traffic between hosts with provided IP range(1) , but rogue insider craftily sends traffic to Switch 1 which is forwarded to another neighboring switch as per rule (2). The Switch 2 accepts communication from any mac address range to destination mac as per rule (3). Hence malicious insider can use indirect policy violation to bypass SLA.

\textbf{Compromised Switch Attack:} If the Switch 1 Cisco-OF-Switch is a fake OpenFlow device (decoy of real OpenFlow switch) in the network, the attacker can present a fake layer 2 port to the victim. The rogue switch can be used to directly insert rule (4) using "ovs-vsctl" commands.  
\begin{equation}
*:00:0a:01|vlan1 \mapsto *:00:0b:03|vlan1 
\end{equation} 
The attacker here establishes a layer 2 tunnel for communication between the source and destination mac address. Multiple opportunistic tunnels of this nature can be established by a rogue switch to achieve resource starvation (DDoS) on the victim machine.

\subsection{Trust Model}
We incorporate the chain of trust model in SDN infrastructure to ensure the protection of entities which are part of the data plane, control plane, and management plane. We have considered hierarchical and decentralized trust framework to secure SDN environment. The Public Key Infrastructure (PKI) based setup will ensure only legitimate devices are allowed to establish data and control plane traffic with each other. \\ 

\subsubsection{Distributed Trust Model}

We use the distributed trust model in scenarios where the management plane application for two security domains in consideration are different, e.g., load balancing application and firewall application. We establish Root-CA to issue local CA privileges to the controller node. The controller node will in turn act as Control-Plane CA-Server. The clients for the control plane will be OpenFlow enabled switches.
\begin{figure}[!ht]
    \centering
    \includegraphics[width=0.5\textwidth]{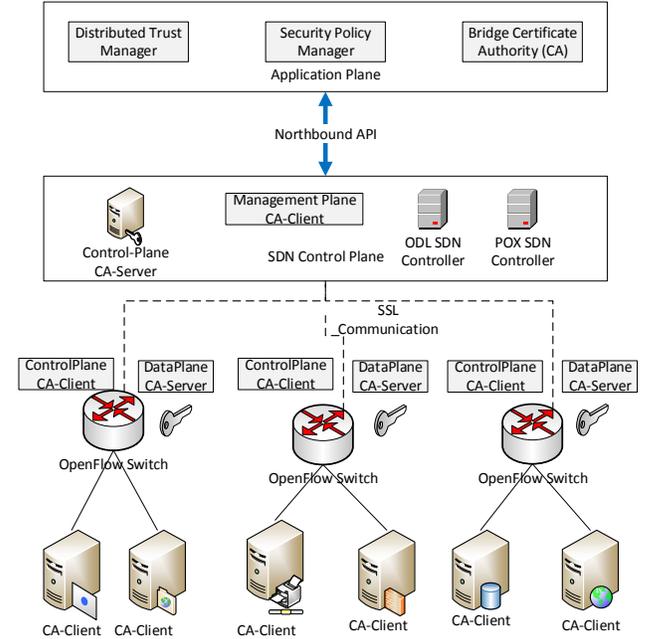}
    \vspace{-6em}
    \caption{Decentralized Trust Model}
    \label{fig:4}
      \vspace{-2em}
\end{figure}

The Data Plane will have individual switches acting as DataPlane CA-Server, as shown in the Figure~\ref{fig:4}. The DataPlane CA-Server (OpenFlow switches) will be issued valid CA certificates by  Control-Plane CA-Server. The DathPlane CA-Server will issue private keys to individual host machines. This type of architecture will have a Bridge-CA which will allow trusted management plane Root-CA to cross-certify each other and in effect achieve a fully distributed trust architecture. The SDN controllers responsible for different network segments and can use Bridge CA for mutually authenticating each other.

\section{Implementation Details}
We used mininet~\cite{mininet} network simulator to test the performance of the algorithm and to create a scalable network environment. We used Opendaylight (ODL) SDN controller running on different ports to conduct experiments. Each ODL controller was responsible for managing one or more security domains. We used a physical server with 16GB RAM and 4-core Intel i7 processor to conduct the performance evaluation. The PKI infrastructure was set up using OpenSSL based APIs to setup Certificate Authority (CA) and issuing private keys to each control and Data Plane elements. The Open vSwitch 2.7.0 with support for DPDK was used. Some customizations to the mininet module to support DPDK. We enabled support for the netdev module in mininet. Poll Mode Driver (PMD) allows parallelization of certificate creation and verification on each host.


\begin{algorithm}
    \caption{SDN-TRUFL Algorithm}\label{euclid}
    \begin{algorithmic}[1]
        
        \Procedure{Setup-CA}{caName}
        \State $\textit{cakey} \gets \textit{createKeyPair(RSA, 2048)}$.
        \State $\textit{careq} \gets \textit{createCertReq(cakey, CN = caName)}$.    
        \EndProcedure
        
        \item[]
        
        \Procedure{keypairSetup}{secDom, controllers}
        
        \For {$\textit{s} \in  \textit{secDom[]}$}
        \State $\textit{thread.create( setuCA(s) , “CA-s”)}$.
        \EndFor
        \For {$\textit{c} \in  \textit{controllers}$}
        \State $\textit{pkey} \gets \textit{c.createKeyPair(RSA, 2048)}$.
        \State $\textit{req}  \gets \textit{createCertReq(pkey, secDom.localCA)}$.
            
        \begin{varwidth}[t]{\linewidth} 
             $\textit{cert} \gets \textit{createCertificate(req, (secDom.localCA}, \\ 
              \textit{cakey),1, (0, 60*60*24*365*5))}$. 
        \end{varwidth}
        \State $\textit{thread.create( setuCA(c) , “CA-c")}$. 
        \EndFor
        
        \For {$\textit{sw} \in  \textit{c}$}
        
        \State $\textit{pkey} \gets \textit{c.createKeyPair(RSA, 2048)}$.       
        \State $\textit{req}  \gets \textit{createCertReq(pkey, C.ControlPlaneCA)}$.
        
        \State \begin{varwidth}[t]{\linewidth} 
            $\textit{cert} \gets \textit{createCertificate(req, (secDom.localCA}, \\ 
            \textit{cakey),1, (0, 60*60*24*365*5))}$.
        \end{varwidth}
        
        \EndFor
        
        \EndProcedure
        
        \item[]
        
        \Procedure{VerifyTrust}{net}
        \For {$\textit{link} \in  \textit{net.links}$}
        
        \State $\textit{pkey} \gets \textit{open(PKI/link.intf1.pkey, 'r').read()}$
        \State $\textit{cert} \gets \textit{open(PKI/link.intf1.cert, 'r').read()}$   
        \State $\textit{pkey-cert} \gets \textit{crypto.load\_privatekey(pkey)}$
        \State $\textit{dsign} \gets \textit{ crypto.sign(pkey-crypto , sha256)}$
        \State $\textit{crypto.verify(pkey-cert, dsign, sha256)}$  
        
        \EndFor
        
        \EndProcedure

    \end{algorithmic}
\end{algorithm}

Algorithm~\ref{euclid}, SDN-TRUFL, is used to establish and verify trust between management plane applications, control plane applications, and data plane elements. The root CA setup lines \textit{1-3} sets up a CA server for the management plane. All the controllers serve as clients to this CA server. Each controller is connected to one or more security domains. Lines \textit{7-10} shows the creation of the key pair (private, public keys) and certificate for controllers.  We have setup duration of the certificate to 5 years in the current algorithm as shown in line 10. The SDN controller acts as 'localCA' for control plane and issues certificates and key pairs to switches as shown in lines \textit{11-14}. We utilize parallelism to setup local CA's as shown in line \textit{11}. The trustfulness of any SDN element can be verified using a digital signature of network elements for each security domain and their respective certificates. The lines \textit{15-21} show the entire network can be probed for trust using public keys and certificates to periodically verify the trust.

\subsection{Distributed Packet Processing in Open vSwitch}
We used poll mode driver (PMD) threads for each connected host NIC separately. Each Open vSwitch was modified after mininet topology creation
to support \textit{netdev} as shown for switch \textit{s1}  with port \textit{s1-eth1} below. \\
\begin{verbatim}
ovs-vsctl add-br s1  set bridge s1 
datapath_type=netdev
ovs-vsctl add-port s1 myportnameone
set Interface myportnameone type=dpdk
\end{verbatim}
Additionally hosts NIC were assigned two transmission and receiver buffers to speed up packet processing and reduce latency. 
\begin{verbatim}
 ovs-vsctl set Interface s1-eth1 
  options:n_rxq_desc=2
 ovs-vsctl set Interface s1-eth2 
 options:n_txq_desc=2
\end{verbatim}

\section{Evaluation Results}
We conducted experiments to measure delay encountered by hosts which are part of the network in packet transfer. The network consisted of two security domains, each controlled by one controller spanned across two virtual servers running on a single physical server. Each virtual server had a separate mininet based security domain. The controllers used SSL communication to issue certificates and verify host machines which are part of Data Plane.

\subsection{TRUFL Trust Verification Latency with Number of Hosts}

\pgfplotstableread{
    0 0.20 0.59 0.43  
    1 1.53 3.18 1.74
    2 1.106 10.84 3.53
    3 5.41 46.44 15.45
}\dataset
\begin{figure}
    \centering
    \begin{tikzpicture}
    \begin{axis}[ybar,
    width=.48\textwidth,
    ymin=0,
    ymax=64,
    xlabel = Number of hosts,      
    ylabel={Latency (ms)},
    xtick=data,
    xticklabels = {
        4 hosts,
        16 hosts,
        64 hosts,
        256 hosts
    },
    major x tick style = {opacity=0},
    minor x tick num = 1,
    minor tick length=1.8ex,
    ]
    \addplot[draw=blue,fill=black!20] table[x index=0,y index=1] \dataset; 
    \addplot[draw=red,fill=black!40] table[x index=0,y index=2] \dataset; 
    \addplot[draw=green,fill=black!60] table[x index=0,y index=3] \dataset; 
    \legend{SDN-No-Trust, Centralized-Trust, SDN-TRUFL}
    \end{axis}
    
    \end{tikzpicture}
    
    \caption{Delay Tolerance in SDN-TRUFL Distributed Framework}
    \label{fig:deltol}
\end{figure}
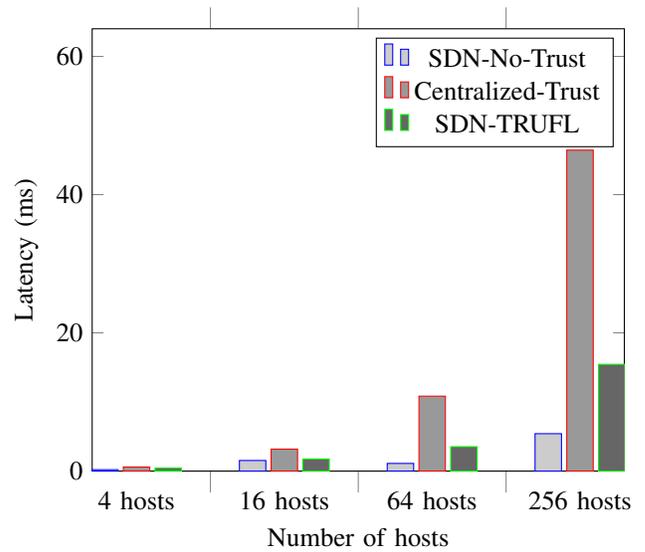        

We calculated round-trip time delay for all hosts when communicating with each other in a fully connected hierarchical PKI network of 4 OpenFlow switches. The root CA will verify local CA's in this setup.  The time required for SDN-TRUFL setup based with DPDK enabled and verification for 16 hosts 3.18 ms compared to normal OpenFlow SDN setup where it is 1.53 ms as shown in Figure~\ref{fig:deltol}. The latency involved in setup and verification of Centralized-SDN framework in a network of 64 hosts is 10.84 ms and latency for a network of 256 hosts is 46.44 ms. The solution scales poorly as can be seen using a direct adaptation of SDN-TRUFL infrastructure. 

We overcame this scalability limitation using a parallel approach using distributed packet processing on each Poll Mode Driver (PMD). Each security domain was assigned a separate thread for the creation of private keys and certificates of each host and switches which are part of its security domain. We observed that latency using Parallel SDN-TRUFL is around 1.74 ms for 16 hosts, 3.53 ms for the 64 hosts and 15.45 ms for 256 host network after using parallelism. Although the latency is still higher than normal OpenFlow network, we achieved better scalability as compared to \textit{Centralized-Trust} implementation using single threaded computation. 

\subsection{End to End Reachability Latency Analysis}
We compared the performance in terms of runtime latency of our research work TRUFL vs Number of flow rules with existing research works that focus on security analysis in an SDN environment - NetSyn~\cite{mcclurg2015efficient}, SDPA~\cite{zhu2015sdpa} or checking network properties such as forwarding latency - Hassel~\cite{kazemian2012header}. 

\begin{table}[htb]
    \centering
    \caption{Forwarding Latency for  TRUFL, Net-Syn~\cite{mcclurg2015efficient}, SDPA~\cite{zhu2015sdpa} and Hassel~\cite{kazemian2012header}}
    \label{tab:1}
    \begin{tabular}{p{1.3cm}|p{0.8cm}|p{1.1cm}|p{0.8cm}|p{0.8cm}}
        
        \hline  
        \textbf{Number of Rules} & \textbf{TRUFL (ms)} & \textbf{Net-Syn (ms)} & \textbf{SDPA (ms)} & \textbf{Hassel (ms)} \\
        \hline
        10,000 & 7 & 25 & 130-140 & 100 \\
        \hline
        20,000 & 11 & 34 & 130- 145 & 1100 \\
        \hline
        30,000 & 14 & 43 & 130-145 & 3000 \\
        \hline
        40,000 & 19  & 57 & 130-145 & 5000 \\
        \hline
        50,000 & 28 & 65 &130-145  &  6000 \\
        \hline
    \end{tabular}                
\end{table}
We checked the end-to-end reachability, which was a common network property for the related work we examined. As can be seen in the Table~\ref{tab:1}, the number of OpenFlow rules increase from 10k to 50k, the latency for checking forwarding also increases for TRUFL from 7 ms to 28 ms. The latency for 50k rules for Net-Syn was 65 ms, whereas for SDPA was in the range 130-145 ms, the exact latency value deduction for SDPA was difficult based on examination of their experimental results. The latency value for Hassel is around 6s (6000ms) for 50k rules. The faster verification of reachability is faster in TRUFL since the use of distributed trust management  with DPDK speeds up the packet processing, hence verification is in the order of few milliseconds. 

\section{Conclusion and Future Work}
The SDN infrastructure provides a great way of providing network orchestration, management, and security. There are, however, some security threat vectors that affect SDN infrastructure and network traffic management. We have identified some important threat vectors as part of this work, including rogue hosts mounting DDoS attacks, malicious switches and controllers, and incorrect traffic flow rules. We proposed TRUFL framework to provide distributed trust management. Since public key cryptography based trust management is slow, we utilized parallel computing to speed up the setup and trust verification. The distributed trust management allowed fast establishment and verification of trust in SDN environment. TRUFL scales well with the increased number of OpenFlow rules in an SDN environment compared to existing works - NetSyn~\cite{mcclurg2015efficient}, SDPA~\cite{zhu2015sdpa}, Hassel~\cite{kazemian2012header} when checking end-to-end packet forwarding latency. We plan to extend this work and include Group Based Policy (GBP) mechanism to establish security policies across multiple tenants. The DPDK based userspace packet processing can be combined with GBP mechanism to allow modular security architecture in the SDN environment.  

\section*{Acknowledgment}

All authors are thankful for research grants from Naval Research Lab N00173-15-G017,  N0017319-1-G002 and National Science Foundation US DGE-1723440, OAC-1642031, SaTC-1528099.
\bibliographystyle{IEEEtran}
\bibliography{template}

\begin{thebibliography}{10}
\providecommand{\url}[1]{#1}
\csname url@samestyle\endcsname
\providecommand{\newblock}{\relax}
\providecommand{\bibinfo}[2]{#2}
\providecommand{\BIBentrySTDinterwordspacing}{\spaceskip=0pt\relax}
\providecommand{\BIBentryALTinterwordstretchfactor}{4}
\providecommand{\BIBentryALTinterwordspacing}{\spaceskip=\fontdimen2\font plus
\BIBentryALTinterwordstretchfactor\fontdimen3\font minus
  \fontdimen4\font\relax}
\providecommand{\BIBforeignlanguage}[2]{{%
\expandafter\ifx\csname l@#1\endcsname\relax
\typeout{** WARNING: IEEEtran.bst: No hyphenation pattern has been}%
\typeout{** loaded for the language `#1'. Using the pattern for}%
\typeout{** the default language instead.}%
\else
\language=\csname l@#1\endcsname
\fi
#2}}
\providecommand{\BIBdecl}{\relax}
\BIBdecl

\bibitem{lim2015security}
A.~Lim, ``Security risks in sdn and other new software issues.''

\bibitem{sdnpki}
\BIBentryALTinterwordspacing
``sdnpki,
  \url{https://www.ietf.org/proceedings/93/slides/slides-93-sdnrg-3.pdf},''
  2015. [Online]. Available:
  \url{https://www.ietf.org/proceedings/93/slides/slides-93-sdnrg-3.pdf}
\BIBentrySTDinterwordspacing

\bibitem{dpdk}
\BIBentryALTinterwordspacing
``Data path development kit (dpdk),'' 2017. [Online]. Available:
  \url{https://software.intel.com/en-us/articles/open-vswitch-with-dpdk-overview}
\BIBentrySTDinterwordspacing

\bibitem{inteldpdk}
\BIBentryALTinterwordspacing
``inteldpdk,
  \url{https://software.intel.com/en-us/articles/set-up-open-vswitch-with-dpdk-on-ubuntu-server},''
  2015. [Online]. Available:
  \url{https://software.intel.com/en-us/articles/set-up-open-vswitch-with-dpdk-on-ubuntu-server}
\BIBentrySTDinterwordspacing

\bibitem{mcclurg2015efficient}
J.~McClurg, H.~Hojjat, P.~{\v{C}}ern{\`y}, and N.~Foster, ``Efficient synthesis
  of network updates,'' in \emph{ACM SIGPLAN Notices}, vol.~50, no.~6.\hskip
  1em plus 0.5em minus 0.4em\relax ACM, 2015, pp. 196--207.

\bibitem{zhu2015sdpa}
S.~Zhu, J.~Bi, C.~Sun, C.~Wu, and H.~Hu, ``Sdpa: Enhancing stateful forwarding
  for software-defined networking,'' in \emph{Network Protocols (ICNP), 2015
  IEEE 23rd International Conference on}.\hskip 1em plus 0.5em minus
  0.4em\relax IEEE, 2015, pp. 323--333.

\bibitem{kazemian2012header}
P.~Kazemian, G.~Varghese, and N.~McKeown, ``Header space analysis: Static
  checking for networks.'' in \emph{NSDI}, vol.~12, 2012, pp. 113--126.

\bibitem{6wind}
\BIBentryALTinterwordspacing
``6wind support for data path development kit (dpdk),'' 2017. [Online].
  Available: \url{https://www.intel.com/content/dam/www/public/us/en/documents
  /presentation/6wind-support-intel-dpdk-presentation.pdf}
\BIBentrySTDinterwordspacing

\bibitem{opensslpki}
\BIBentryALTinterwordspacing
``opoensslpki,
  \url{https://media.readthedocs.org/pdf/pki-tutorial/latest/pki-tutorial.pdf},''
  2017. [Online]. Available:
  \url{https://media.readthedocs.org/pdf/pki-tutorial/latest/pki-tutorial.pdf}
\BIBentrySTDinterwordspacing

\bibitem{porras2012security}
P.~Porras, S.~Shin, V.~Yegneswaran, M.~Fong, M.~Tyson, and G.~Gu, ``A security
  enforcement kernel for openflow networks,'' in \emph{Proceedings of the first
  workshop on Hot topics in software defined networks}.\hskip 1em plus 0.5em
  minus 0.4em\relax ACM, 2012, pp. 121--126.

\bibitem{kreutz2013towards}
D.~Kreutz, F.~Ramos, and P.~Verissimo, ``Towards secure and dependable
  software-defined networks,'' in \emph{Proceedings of the second ACM SIGCOMM
  workshop on Hot topics in software defined networking}.\hskip 1em plus 0.5em
  minus 0.4em\relax ACM, 2013, pp. 55--60.

\bibitem{koponen2010onix}
T.~Koponen, M.~Casado, N.~Gude, J.~Stribling, L.~Poutievski, M.~Zhu,
  R.~Ramanathan, Y.~Iwata, H.~Inoue, T.~Hama \emph{et~al.}, ``Onix: A
  distributed control platform for large-scale production networks.'' in
  \emph{OSDI}, vol.~10, 2010, pp. 1--6.

\bibitem{tootoonchian2010hyperflow}
A.~Tootoonchian and Y.~Ganjali, ``Hyperflow: A distributed control plane for
  openflow,'' in \emph{Proceedings of the 2010 internet network management
  conference on Research on enterprise networking}, 2010, pp. 3--3.

\bibitem{chowdhary2016sdn}
A.~Chowdhary, S.~Pisharody, and D.~Huang, ``Sdn based scalable mtd solution in
  cloud network,'' in \emph{Proceedings of the 2016 ACM Workshop on Moving
  Target Defense}.\hskip 1em plus 0.5em minus 0.4em\relax ACM, 2016, pp.
  27--36.

\bibitem{betge2015trust}
S.~Betg{\'e}-Brezetz, G.-B. Kamga, and M.~Tazi, ``Trust support for sdn
  controllers and virtualized network applications,'' in \emph{Network
  Softwarization (NetSoft), 2015 1st IEEE Conference on}.\hskip 1em plus 0.5em
  minus 0.4em\relax IEEE, 2015, pp. 1--5.

\bibitem{phemius2014disco}
K.~Phemius, M.~Bouet, and J.~Leguay, ``Disco: Distributed multi-domain sdn
  controllers,'' in \emph{Network Operations and Management Symposium (NOMS),
  2014 IEEE}.\hskip 1em plus 0.5em minus 0.4em\relax IEEE, 2014, pp. 1--4.

\bibitem{shin2013fresco}
S.~Shin, P.~A. Porras, V.~Yegneswaran, M.~W. Fong, G.~Gu, and M.~Tyson,
  ``Fresco: Modular composable security services for software-defined
  networks.'' 2013.

\bibitem{pisharody2016conflicts}
S.~Pisharody, A.~Chowdhary, and D.~Huang, ``Security policy checking in
  distributed {SDN} based clouds,'' in \emph{2016 IEEE Conference on
  Communications and Network Security (CNS) (IEEE CNS 2016)}, Oct. 2016.

\bibitem{huang2018software}
D.~Huang, A.~Chowdhary, and S.~Pisharody, \emph{Software-Defined Networking and
  Security: From Theory to Practice}.\hskip 1em plus 0.5em minus 0.4em\relax
  CRC Press, 2018.

\bibitem{cheng2011rendezvous}
N.~Cheng, K.~Govindan, and P.~Mohapatra, ``Rendezvous based trust propagation
  to enhance distributed network security,'' \emph{International Journal of
  Security and Networks}, vol.~6, no. 2-3, pp. 112--122, 2011.

\bibitem{mininet}
\BIBentryALTinterwordspacing
``Mininet virtual network,'' 2015. [Online]. Available:
  \url{http://mininet.org/}
\BIBentrySTDinterwordspacing

\end{thebibliography}

\end{document}